\documentclass[aps,twocolumn,preprintnumbers,superscriptaddress,prx]{revtex4}
\usepackage{graphicx}
\usepackage{amsmath}
\usepackage{amssymb}
\usepackage{color}

\begin{document}

\title{Laser Mode Bifurcations Induced by $\mathcal{PT}$-Breaking Exceptional Points}

\author{Bofeng Zhu}

\affiliation{Division of Physics and Applied Physics, School of Physical and Mathematical Sciences,\\
Nanyang Technological University, Singapore 637371, Singapore}

\author{Qi Jie Wang}

\affiliation{School of Electrical and Electronic Engineering, \\
  Nanyang Technological University, Singapore 637371, Singapore}

\author{Y.~D.~Chong}

\affiliation{Division of Physics and Applied Physics, School of Physical and Mathematical Sciences,\\
Nanyang Technological University, Singapore 637371, Singapore}

\affiliation{Centre for Disruptive Photonic Technologies, Nanyang Technological University, Singapore 637371, Singapore}

\email{yidong@ntu.edu.sg}

\begin{abstract}
  A laser consisting of two independently-pumped resonators can
  exhibit mode bifurcations that evolve out of the exceptional
  points (EPs) of the linear system at threshold.  The EPs are
  non-Hermitian degeneracies occurring at the parity/time-reversal
  ($\mathcal{PT}$) symmetry breaking points of the threshold system.
  Above threshold, the EPs become bifurcations of the nonlinear
  zero-detuned laser modes, which can be most easily observed by
  making the gain saturation intensities in the two resonators
  substantially different.  Small pump variations can then switch
  abruptly between different laser behaviors, e.g.~between
  below-threshold and $\mathcal{PT}$-broken single-mode operation.
\end{abstract}

\maketitle

\textit{Introduction.}---Non-Hermitian effects have been the subject of strong and sustained attention in optics and related fields, driven largely by the rise of parity/time-reversal ($\mathcal{PT}$) symmetric photonics \cite{Longhi2017,feng2017,El-Ganainy2017}.  Following early proposals to utilize $\mathcal{PT}$ symmetry for unidirectional invisibility cloaks \cite{lin2011,mostafazadeh2013}, laser-absorbers \cite{Longhi2010, Chong2011, Wong2016}, etc., much recent progress has been based on the complex interactions between $\mathcal{PT}$ symmetry breaking and nonlinearity \cite{Musslimani2008, Lumer2013,Hassan2015,konotop2016,suchkov2016}.  Recent prominent experiments exploring this direction have demonstrated efficient optical isolation \cite{peng2014ep, chang2014, Zhou2016}, robust wireless power transfer \cite{Assawaworrarit2017}, and the stabilization of single-mode lasing in microcavity lasers \cite{Miri2012, Hodaei2014, Feng2014, Teimourpour2017}.

Lasers provide a compelling setting for such investigations, as they are intrinsically both non-Hermitian (due to gain and outcoupling loss), and nonlinear when operating above threshold (due to gain saturation).  Although these features have long been known, the recent interest in non-Hermitian physics has provided fresh inspiration for devising lasers with novel characteristics.  Several authors have drawn special attention to the peculiar effects of exceptional points (EPs)---points in parameter space where a non-Hermitian Hamiltonian becomes defective and two (or more) eigenvectors coalesce \cite{heiss2004, berry2004}.  $\mathcal{PT}$ symmetry provides a convenient, though not exclusive, way to generate EPs: whenever $\mathcal{PT}$ symmetry spontaneously breaks, there is an EP at the transition point \cite{Ramezani2012}.  The presence of an EP among the non-Hermitian eigenmodes of a laser has been shown to give rise to pump-induced suppression and revival of lasing \cite{Liertzer2012, Brandstetter2014, Peng2014}, though these works regarded the EP as ``accidental'', not arising from $\mathcal{PT}$ symmetry breaking.  EPs have also been shown to promote single-mode operation in dark state lasers \cite{gentry2014, Hodaei2016} and $\mathcal{PT}$-symmetric lasers \cite{Miri2012, Hodaei2015}.

In this paper, we show that EPs associated with $\mathcal{PT}$ symmetry breaking in a laser at threshold can generate bifurcations in the nonlinear laser modes above threshold. The bifurcations appear as discontinuities in the laser's I-V curve (i.e., the dependence of output power on pump strength), allowing small variations in the pump strength to induce abrupt switching between below-threshold and single-mode operation, or between high-power $\mathcal{PT}$-broken single-mode operation and low-power $\mathcal{PT}$-symmetric two-mode operation.

Bifurcations have previously been investigated extensively in laser physics; they underlie the operation of bistable lasers, which have applications as optical ``flip-flop'' memory devices \cite{lamb1964, lasher1964, Harder1981, Chen1987, Chen1987_2, Tang1987, Kawaguchi1992, Johnson1993,  kawaguchi1997, hill2004, ma2017}.  Typically, they result from the inclusion of two different nonlinearities in a single laser, such as a saturable gain medium and a saturable absorber \cite{lasher1964, Harder1981, hill2004, ma2017}, or two polarization modes experiencing different gain saturation \cite{Chen1987, Chen1987_2, Kawaguchi1992, kawaguchi1997}.

In this context, the most noteworthy aspect of the present work is that it establishes a connection between the phenomenon of laser mode bifurcations and the physics of $\mathcal{PT}$ symmetry and EPs.  We study an exemplary coupled-resonator system in which the two resonators can be pumped independently, and show that laser mode bifurcations evolve continuously out of the EPs corresponding to $\mathcal{PT}$ symmetry breaking points in the family of $\mathcal{PT}$-symmetric threshold modes.  These EPs are also closely related to the phenomenon of pump-induced suppression and revival of lasing studied in Refs.~\onlinecite{Liertzer2012, Brandstetter2014, Peng2014}.  When nonlinear effects are neglected, varying the pump on one of the resonators causes the system to evolve between $\mathcal{PT}$-symmetric and $\mathcal{PT}$-broken laser modes, with lasing suppressed in an interval between the two regimes (in the vicinity of the EP).  When the nonlinearities are included, however, imbalances between the gain saturation intensities of the two resonators shift the domain of stability of the $\mathcal{PT}$-broken mode, producing a bifurcation.  Our analysis is based on coupled mode theory \cite{CMT}, via analysis of the steady-state solutions backed by time-domain simulations.

\begin{figure}
  \centering
  \includegraphics[width=0.48\textwidth]{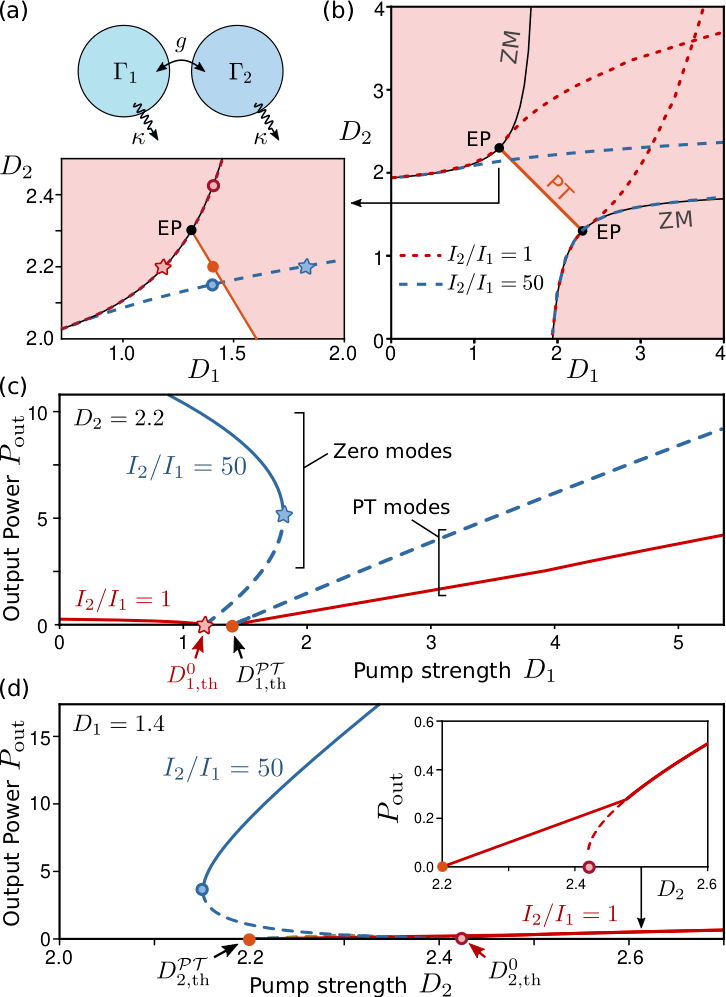}
  \caption{(a) Schematic of coupled-resonator laser.  (b) Lasing threshold conditions for pump strengths $D_1$ and $D_2$.  The $\mathcal{PT}$-symmetric threshold (orange, labelled ``PT'') and zero-mode threshold (black, labelled ``ZM'') are plotted as solid lines.  Red areas show where the linear Hamiltonian has one or more amplifying eigenvalues, the conventional criterion for lasing.  Dashes indicate the zero-mode bifurcation points for different saturation intensity ratios $I_2/I_1$.  Left panel: the vicinity of the linear Hamiltonian's exceptional point (EP), with stars and circles showing the pump values corresponding to the same symbols in (c) and (d).  (c) Output power $P_{\mathrm{out}}$ versus $D_1$, for $D_2 = 2.2$.  For $I_1 = I_2$ (red), the laser exhibits the ``suppression and revival'' phenomenon~\cite{Liertzer2012, Brandstetter2014, Peng2014}.  For $I_2/I_1 = 50$, there is a high-intensity stable lasing zero mode (blue solid line) ending in a bifurcation; the other modes (blue dashes) are unstable.  (d) $P_{\mathrm{out}}$ versus $D_2$ for $D_1 = 1.4$, showing both stable modes (solid lines) and unstable modes (dashes).  For $I_2/I_1 = 50$, a stable mode appears abruptly at finite intensity (blue circle).  The other model parameters, for all subplots, are $\kappa = 1.8$, $g = 0.5$, and $I_1 = 1$.}
  \label{fig:thresholds}
\end{figure}

\textit{Dimer laser model.}---Consider a pair of coupled resonators, as shown in Fig.~\ref{fig:thresholds}(a).  They have the same resonance frequency, variable gain rates $\Gamma_1$ and $\Gamma_2$, equal radiative loss rates $\kappa$, and coupling rate $g$, with $\Gamma_1$, $\Gamma_2$, $\kappa$, and $g$ all real.  In the framework of coupled-mode theory \cite{CMT}, a single-mode steady-state solution can be described by the equation
\begin{equation}
  \begin{pmatrix}
    i(\Gamma_1-\kappa)&g\\g&i(\Gamma_2-\kappa)
  \end{pmatrix}
  \begin{pmatrix}
    \Psi_1\\\Psi_2
  \end{pmatrix}
  = \Omega
  \begin{pmatrix}
    \Psi_1\\\Psi_2
  \end{pmatrix},
  \label{cmt_equation}
\end{equation}
where $\Psi_1, \Psi_2$ are the complex field amplitudes in each resonator, and $\Omega$ is the relative frequency.  We assume the field amplitudes are the only relevant dynamical variables (i.e., the polarization and population inversion have no independent dynamics, which is usually the case in semiconductor lasers \cite{Haken}).  In each resonator $j$, the gain has the saturable form $\Gamma_j = D_j/(1+|\Psi_j|^2/I_j)$, where $D_j \ge 0$ is the pump strength and $I_j > 0$ is the saturation intensity.  Laser modes are solutions to Eq.~\eqref{cmt_equation} with $|\Psi_1|, |\Psi_2| > 0$ and $\Omega \in \mathbb{R}$, while a threshold mode has real $\Omega$ with linear gain $\Gamma_j = D_j$ (i.e., zero intensity).  Evidently, the threshold conditions are independent of $I_1$ and $I_2$.  While multiple threshold modes may exist, lasers typically have a ``lowest'' threshold featuring the smallest possible pump(s); with increasing pump, a stable laser mode evolves from the lowest threshold mode, with output power increasing continuously from zero.  We shall see, however, that the dimer can behave very differently.

Eq.~\eqref{cmt_equation} gives two types of threshold modes. Firstly, if $\Omega \ne 0$, the real and imaginary parts of the secular equation give $D_1+D_2 = 2\kappa$ and $\Omega^2 =g^2-(D_2-\kappa)^2$.  For $\Omega$ to be real, we also require
\begin{equation}
  \kappa-g \le D_1, D_2 \le \kappa + g.
  \label{omega_real}
\end{equation}
We call such solutions ``$\mathcal{PT}$-symmetric threshold modes'', for they correspond to the $\mathcal{PT}$-unbroken eigenvectors of a Hamiltonian whose imaginary diagonal components are equal and opposite.

The second type of threshold mode occurs when $\Omega = 0$.  We call these ``zero modes'' for short.  In this case, the secular equation gives $(D_1-\kappa)(D_2-\kappa) + g^2 = 0$, and the Hamiltonian is not $\mathcal{PT}$ symmetric.  This is \textit{not}, however, an ordinary instance of linear $\mathcal{PT}$ symmetry breaking, wherein a linear Hamiltonian remains $\mathcal{PT}$-symmetric but the eigenmodes become $\mathcal{PT}$-broken \cite{Longhi2017,feng2017,El-Ganainy2017}.  Here, the Hamiltonian itself loses $\mathcal{PT}$ symmetry.

The two threshold conditions are plotted in Fig.~\ref{fig:thresholds}(b) against the independent pump strengths $D_1$ and $D_2$.  The $\mathcal{PT}$-symmetric thresholds lie on a line segment whose end-points are exceptional points (EPs) where the $\mathcal{PT}$-symmetric Hamiltonian becomes defective, and the inequalities in \eqref{omega_real} saturate.  These EPs are also where $\mathcal{PT}$-symmetric and zero-mode threshold curves meet.  One would ordinarily expect the dimer to be below threshold in the white region of Fig.~\ref{fig:thresholds}(b), bounded by the threshold curves.  In the red region, at least one eigenfrequency is amplifying [$\mathrm{Im}(\Omega) > 0$] in the zero intensity limit; thus, when entering this region starting from one of the threshold curves, one expects the output power to increase continuously from zero.  An extremely similar threshold map has previously been observed experimentally in Ref.~\onlinecite{Brandstetter2014}, though the relation to $\mathcal{PT}$ symmetry was not discussed in that work.

Fig.~\ref{fig:thresholds}(c) plots the output power $P_{\mathrm{out}}$ versus pump strength $D_1$, where the output power is given by $P_{\mathrm{out}} = \kappa \left( |\psi_1|^2 + |\psi_2|^2\right)$.  For fixed $D_2 > 0$ and $I_1 = I_2$ (red curves), the system lases in a zero mode for small $D_1$; then as $D_1$ increases, $P_{\mathrm{out}}$ drops to zero over a finite range of $D_1$, then increases as the system resumes lasing in a $\mathcal{PT}$-symmetric mode.  This is the phenomenon of ``suppression and revival of lasing''~\cite{Liertzer2012, Brandstetter2014, Peng2014}. The role of nonlinearity in transitioning from a non-$\mathcal{PT}$-symmetric mode to $\mathcal{PT}$-symmetric mode has also been noted by Hassan \textit{et al.}~\cite{Hassan2015}.

\begin{figure}
  \centering
  \includegraphics[width=0.95\linewidth]{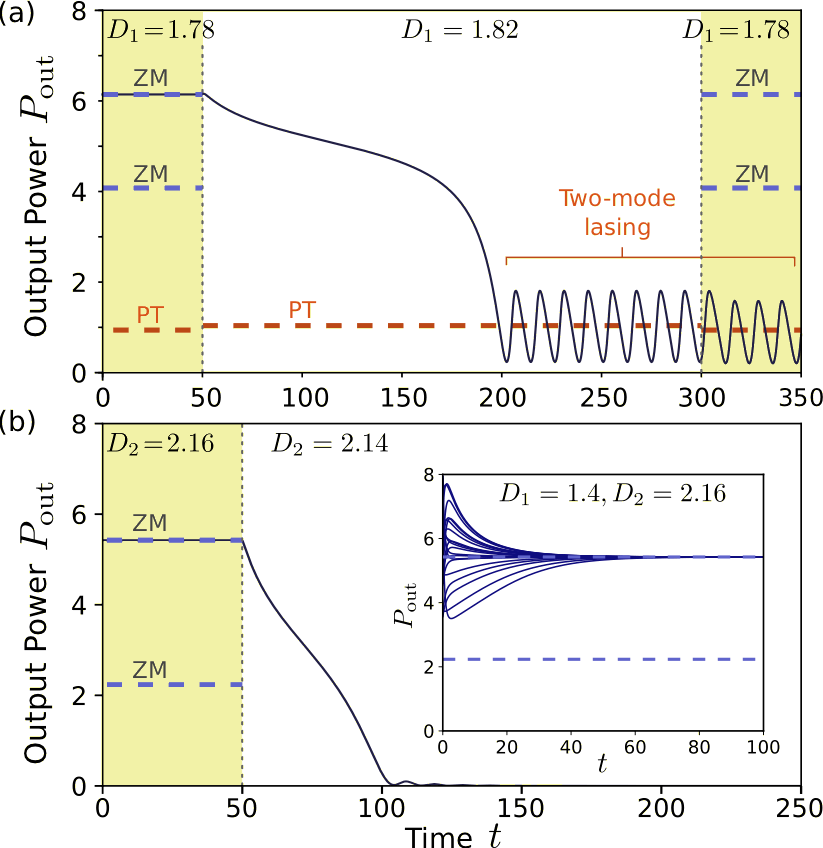}
  \caption{Simulations showing the time evolution of the output power
    $P_{\mathrm{out}}$ as the pump is tuned across the bifurcation
    point.  In all subplots, $\kappa = 1.8$, $g = 0.5$, $I_1 = 1$, and
    $I_2 = 50$.  Vertical dotted lines indicate when the change in
    pump occurs, and horizontal dashes indicate the steady-state
    zero-mode and $\mathcal{PT}$-symmetric mode intensities before and
    after the change.  (a) With fixed $D_2 = 2.2$, $D_1$ is increased
    from $1.78$ to $1.82$ at $t = 50$, so as to cross the bifurcation
    in Fig.~\ref{fig:thresholds}(c) left-to-right, returning to $D_1 =
    1.78$ at $t = 300$.  The system, initialized in the
    higher-intensity zero mode, switches to two-mode lasing with
    $P_{\mathrm{out}}$ beating around the $\mathcal{PT}$-symmetric
    modal intensity at $t = 50$.  It remains there after $t = 300$,
    thus showing flip-flop behavior.  (b) With fixed $D_1 = 1.4$,
    $D_2$ is decreased from $2.16$ to $2.14$, crossing the bifurcation
    in Fig.~\ref{fig:thresholds}(d) right-to-left.  The system is
    initialized in the higher-intensity zero mode, and decays to zero
    intensity.  Inset: time-evolution for $D_1 = 1.4$ and $D_2 =
    2.16$, with noisy initial conditions to test the zero mode's
    stability.  The different solid curves show results for 20
    independent initial conditions $\Psi_j(0) = \Psi_j^{\mathrm{ZM}} +
    \delta \Psi_j$, where $\Psi_j^{\mathrm{ZM}}$ is the zero-mode
    solution, $\delta \Psi_j \sim 0.2 [\mathcal{N} + i\mathcal{N}]$,
    and $\mathcal{N}$ is the standard normal distribution.  }
  \label{fig:timedomain}
\end{figure}

When the gain media have substantially different saturation intensities, we find that the behavior of the nonlinear system deviates from the above simple predictions based on the threshold map.  As shown by the blue curves in Fig.~\ref{fig:thresholds}(c), for $I_2/I_1 = 50$ the lasing zero-mode's power decreases with $D_1$, but does \textit{not} reach zero at the zero-mode threshold point $D_{1,\mathrm{th}}^0$.  Instead, the solution branch ends in a bifurcation point, at nonzero intensity and at a value of $D_1$ higher than both $D_{1,\mathrm{th}}^0$ and the $\mathcal{PT}$-symmetric threshold $D_{1,\mathrm{th}}^{\mathcal{PT}}$.  In effect, the domain of nonlinear zero-mode solutions shifts beyond the boundaries set by the threshold conditions, as indicated by the dashed curves in Fig.~\ref{fig:thresholds}(b).  Lyanpunov stability analysis shows that the entire high-intensity branch of lasing zero-modes is stable up to the bifurcation point.  A second branch of lower-intensity zero-modes, starting from $D_{1,\mathrm{th}}^0$ and ending at the bifurcation point, is unstable, whereas the $\mathcal{PT}$-symmetric modes becomes unstable in single-mode operation shortly above $D_{1,\mathrm{th}}^{\mathcal{PT}}$.  In Fig.~\ref{fig:timedomain}(a), we verify these features via a simulation based on the time-domain version of Eq.~\eqref{cmt_equation}.  As $D_1$ is increased by a small amount (at $t = 50$), crossing the bifurcation point, the laser switches from high-intensity zero-mode operation into two-mode $\mathcal{PT}$-symmetric lasing, which manifests in the time-domain results as a beating around a much lower mean intensity.  When we subsequently set $D_1 = 1.78$ at $t = 300$, the laser retains the two-mode $\mathcal{PT}$-symmetric operation, consistent with the flip-flop behavior observed in previous bistable laser devices.

It is also possible to switch directly between stable laser operation at a finite intensity and the below-threshold regime.  As indicated in Fig.~\ref{fig:thresholds}(b), this happens in the vicinity of one of the EPs, where the domain of stability for the nonlinear zero-modes extends into the ``below-threshold'' part of the threshold map.  The phenomenon is most apparent when the two saturation intensities are very different. In the I-V plot of Fig.~\ref{fig:thresholds}(d), we see that for $I_2/I_1 = 50$, an infinitesimal change in $D_2$ across the bifurcation point (with $D_1$ fixed) causes $P_{\mathrm{out}}$ to jump between zero and a nonzero value.  This is verified by the time-domain simulation results shown in Fig.~\ref{fig:timedomain}(b).  Here, the laser initially operates at the high-intensity zero mode, and after $D_2$ is decreased by less than 1 percent, the intensity drops to zero.  One important limitation to this switching behavior is that the time scale over which switching occurs becomes longer as one approaches the bifurcation point.  When operating very near to the bifurcation point, the present coupled-mode treatment may no longer be sufficient, as other effects (such as gain medium dynamics) could become important.

\begin{figure}
  \centering
  \includegraphics[width=0.95\linewidth]{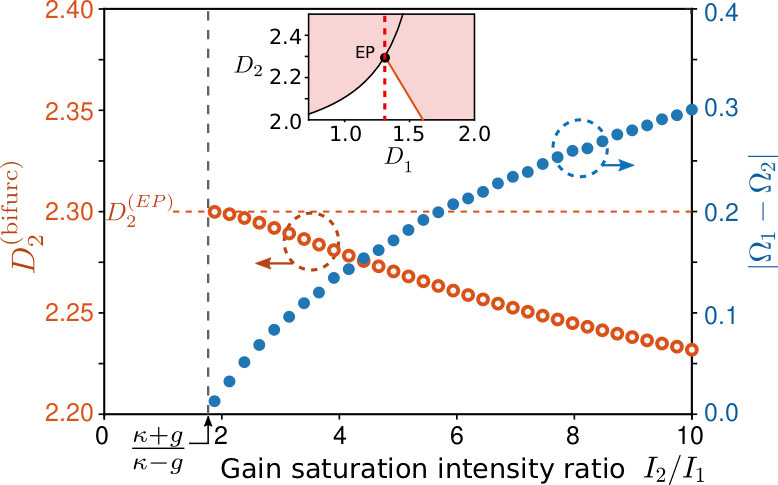}
  \caption{Continuity between the EP of the system at threshold and
    the bifurcation points of the zero modes above threshold.  For
    different values of $I_2/I_1$, the left vertical axis plots the
    value of $D_2$ at the bifurcation point, while the right vertical
    axis plots the magnitude of the difference between the eigenvalues
    of the nonlinear Hamiltonian at the bifurcation point.  The pump
    $D_1$ is fixed at $\kappa-g$, so that the threshold system passes
    through the EP upon varying $D_2$ (threshold diagram shown inset).
    As $I_2/I_1 \rightarrow (\kappa+g)/(\kappa-g)$, the critical value
    of $D_2$ approaches that of the EP, and the eigenvalues coalesce
    as the nonlinear Hamiltonian approaches the EP.  The model
    parameters are $\kappa = 1.8$, $g = 0.5$, and $I_1 = 1$.  }
  \label{fig:bifurcationevo}
\end{figure}

\textit{Relation between exceptional and bifurcation points.}---Generally, the bifurcation points of the nonlinear zero modes are \textit{not} EPs.  At each bifurcation point, the nonlinear Hamiltonian has both a zero eigenvalue, and a nonzero eigenvalue with a distinct eigenvector.  Nonetheless, we can show that the bifurcations evolve continuously out of the EPs of the linear system at threshold.

In Fig.~\ref{fig:bifurcationevo}, we fix $D_1 = \kappa - g$ and locate the value of $D_2$ corresponding to the zero mode bifurcation point, for varying saturation intensity ratio $I_2/I_1$.  The bifurcation is present only for $I_2/I_1$ above a certain minimum value (below this value, the upper zero mode branch is the only one with a valid non-negative intensity).  At the critical value $I_2/I_1 = (\kappa + g)/(\kappa - g)$, the bifurcation point occurs at $D_2 = \kappa + g$, which is the precise location of the EP in the threshold system.  We also calculate the eigenvalues of the nonlinear Hamiltonian at the bifurcation point, $\Omega_{1,2}$, and plot $|\Omega_1 - \Omega_2|$.  As the bifurcation point approaches the EP value, we find that $|\Omega_1 - \Omega_2| \rightarrow 0$; though not plotted in this figure, the overlap between the eigenvectors also approaches unity (i.e., the eigenvectors coalesce), indicating that the bifurcation point indeed becomes coincident with the EP.

\textit{Discussion.}---We have shown that a coupled-resonator laser with independent pumps exhibits mode bifurcations that evolve out of the exceptional points (EPs) of the linear system at threshold.  This is consistent with studies in other systems that have demonstrated that the behavior of nonlinear non-Hermitian systems can be subtly influenced by EPs in the linear limit \cite{Musslimani2008, Lumer2013,Hassan2015,konotop2016,suchkov2016}.  The present case offers the propect of using $\mathcal{PT}$ symmetry principles, specifically the fact that EPs occur at $\mathcal{PT}$-breaking points, to design laser flip-flops and related devices.

A coupled-resonator laser of this type should be straightforward to implement, and indeed a similar design has previously been realized in experimental studies of the ``suppression and revival of lasing'' \cite{Brandstetter2014, Peng2014}.  The key extra requirement, to make the bifurcations easy to observe, is for the resonators to have different gain saturation intensities.  This could be achieved by deliberate engineering of the two gain media.  Alternatively, both resonators can be given identical (but independently-pumped) gain media, but a saturable absorber can be placed on on one of the resonators. Since both gain and absorbing media are unsaturated at threshold, such a configuration only shifts the threshold curves of Fig.~\ref{fig:thresholds} by a fixed amount.  The EPs remain well-defined, and are continuable to the laser mode bifurcations in the same way we have discussed.

\end{document}